\documentclass{epl}

\usepackage{graphicx}
\usepackage{amsmath}
\usepackage{amssymb}

\title{Quasi-1D Bose-Einstein condensates in the dimensional crossover regime}
\author{F.\ Gerbier\footnote{Current address: Institut f{\"u}r Physik, Johannes Gutenberg Universit{\"a}t, 55128 Mainz, Germany; e-mail: gerbier@uni-mainz.de}}
\institute{Laboratoire Charles Fabry de l'Institut
d'Optique\footnote{UMRA 8501 du CNRS}, 91403 Orsay, France}

\begin{document} {Static properties of condensates; thermodynamical, statistical and structural properties}
%\pacs{03.75.Fi}{First pacs description} \pacs{03.75.-b}{Second
%pacs description} \pacs{05.30.Jp}{Third pacs description}

\maketitle

\begin{abstract}
We study theoretically the dimensional crossover from a
three-dimensional elongated condensate to a one-dimensional
condensate as the transverse degrees of freedom get frozen by
tight confinement, in the limit of small density fluctuations,
{\it i.e.} for a strongly degenerate gas. We compute analytically
the radially integrated density profile at low temperatures using
a local density approximation, and study the behavior of phase
fluctuations with the transverse confinement. Previous studies of
phase fluctuations in trapped gases have either focused on the 3D
elongated regimes or on the 1D regime. The present approach
recovers these previous results and is able to interpolate between
them. We show in particular that in this strongly degenerate limit
the shape of the spatial correlation function is insensitive to
the transverse regime of confinement, pointing out to an almost
universal behavior of phase fluctuations in elongated traps.
\end{abstract}

In the recent years, one-dimensional (1D) ultracold atomic gases have been produced
in very elongated magnetic \cite{gorlitz2001a,schreck2001a} or
optical traps \cite{greiner2001a,moritz2003a}, with such tight confinement in two transverse directions that the atomic
motion``freezes'' to radial zero-point oscillations. The equilibrium phase diagram of these trapped 1D gases
shows a rich behavior \cite{petrov2000a}: at low densities the cloud behaves as a gas of impenetrable bosons (``Tonks-Girardeau gas'' \cite{tonks}), and for higher densities (corresponding to most current experimental setups), the cloud is a ``quasi-condensate´´ \cite{petrov2000a,qbec}, characterized by suppressed density fluctuations and the same kind of local correlations as in a true condensate, but also by the lack of long-range phase coherence due to significant phase fluctuations. The latter die out continuously around
a characteristic temperature $T_\phi$, below which the quasicondensate turns into a true condensate. It was realized in \cite{petrov2001a} that such quasicondensates could also exist in very elongated, but
three-dimensional (3D) traps: although atomic motion is possible
in all directions, the lowest-lying excitations of these systems,
which dominate the long range decay of the phase correlations, are
1D in character \cite{stringari1998a}. These 3D quasicondensates
were observed in equilibrium \cite{dettmer2001a} and
non-equilibrium samples \cite{schvarchuck2002a}, and their coherence properties studied through Bragg
spectroscopy \cite{richard2003a,gerbier2003b} or matter-wave
interferometry \cite{hellweg2003a}.

Currently, although the limiting 3D or 1D cases are well
understood theoretically, the kinematic crossover from the ``3D
elongated'' to the truly 1D regime of transverse confinement have
been less studied, despite being relevant to several recent
experiments with ``conventional traps''
\cite{gorlitz2001a,richard2003a}, or to ongoing research on the
manipulation of coherent atomic ensembles on micro-fabricated
substrates \cite{hansel2001a,ott2001a}. For the case of an
infinitely long waveguide with uniform axial density, a many-body
theory has been developed in \cite{das2002a} to describe this
crossover (see also \cite{menotti2002a}). The influence of an
axial trapping potential on the density profile and collective
excitations was addressed in \cite{menotti2002a} by numerical
integration of the transverse dynamics, in \cite{das2002b} through
a separability {\it ansatz}, which is correct in the 1D limit but
not in the 3D case, and in \cite{fuchs2003a}, where solvable
hydrodynamic models that reproduce the 3D and 1D TF limits are
introduced.

In the present paper, we follow a different route and introduce an
accurate analytical approximation for the ``equation of state'' of
the quasi-1D gas, that connects smoothly to the 3D to the 1D
mean-field regimes. In the local density approximation (LDA), we
are then able to work out analytically the 1D (integrated over
radial degrees of freedom) density profile $n_1(z)$, from which
many properties of the trapped quasi-1D gas can be calculated. As
an application, we investigate the influence of the transverse
confinement on phase-fluctuations in a quasi-1D geometry. Ref.
\cite{alkhawaja2004a} considers the related problem at zero
temperature case in a box geometry, and investigate the 3D to 1D
crossover as the box cross-section is reduced. In the trapped case
considered in this paper, quantum fluctuations of the phase are
small away from the Tonks regime \cite{petrov2000a}, so that we
only consider thermal fluctuations. We calculate the spatial
correlation function of the radially averaged atomic field
operator $\hat{\Psi}$,
\begin{equation}
\label{C1} \mathcal{C}^{(1)}(s)= \int d\overline{z} ~\langle
\hat{\Psi}^{\dagger} (\overline{z}+s/2)
\hat{\Psi}(\overline{z}-s/2) \rangle
\end{equation}
at finite (but low) temperatures. This important quantity gives a
global measure of phase coherence across the atomic cloud, and is
also the Fourier transform of the momentum distribution (see for
instance \cite{cctcargese}). The local density approach in this
context is nothing else than a slowly-varying-envelope
approximation, applied to the long-wavelength excitations
responsible for the fluctuations of the phase. We show that this
turns out to be a good approximation even if $T \gtrsim T_\phi$,
provided density fluctuations are small. More importantly, we find
the remarkable property that the shape of the spatial correlation
function (or equivalently the momentum distribution) of the
quasi-condensate is almost insensitive to the precise regime of
transverse confinement, pointing out to the universal character of
phase fluctuations in very elongated traps.

We follow the general method introduced in Ref.
\cite{petrov2000a}, assuming a weakly interacting gas well below
the degeneracy temperature (see \cite{Td} and related discussion
at the end of the paper). Then, density fluctuations are small and
the equilibrium 3D density profile $n_0$ in the trapping potential
$V_{\rm trap}({\bf r})=m\omega_\perp^2\rho^2/2+V(z)$ is given by
the solution of the usual Gross-Pitaevskii equation, even if the
cloud is a quasi-condensate. Without loss of generality, we
introduce the 1D density $n_1(z)=\int d^{(2)}{\bf \rho}~
n_0(\rho,z)$ obtained by integration over the transverse plane,
and the radial mode $f_\perp$ through $n_0(\rho,z)=\mid
f_\perp(\rho,z) \mid^2 n_1(z)$. Assuming a sufficiently shallow
axial confinement and neglecting the density derivatives of $n_1$,
$f_\perp$ with respect to $z$ obtains
\begin{equation} \label{gp2d}
-\frac{\hbar^2}{2M}\frac{\Delta_\perp f_\perp}{f_\perp} +
\frac{1}{2}M\omega_\perp^2\rho^2 + U n_1(z) \mid f_\perp \mid^2=
\mu_{\rm l.e.}[n_1(z)].
\end{equation}
Here, we have used cylindrical coordinates, denoting the
transverse radius as $\rho$ and the longitudinal coordinate as
$z$. The 3D coupling constant $U$ is related to the s-wave
scattering length $a$ through $U=4\pi\hbar^2a/M$, where $M$ is the
atomic mass. The local equilibrium chemical potential $\mu_{\rm
l.e.}$ depends on $z$ through
\begin{equation}\label{lda}
\mu_{\rm l.e.}[n_1(z)]+V(z)=\mu,
\end{equation}
with $\mu$ the global chemical potential of the cloud in the trap. The relation between $n_1$ and
$\mu_{\rm l.e.}$, that includes the effect of transverse
confinement, can be seen as an effective ``equation of state'' for
the 1D gas \cite{menotti2002a}.

\begin{figure}[t]
\includegraphics[width=12cm]{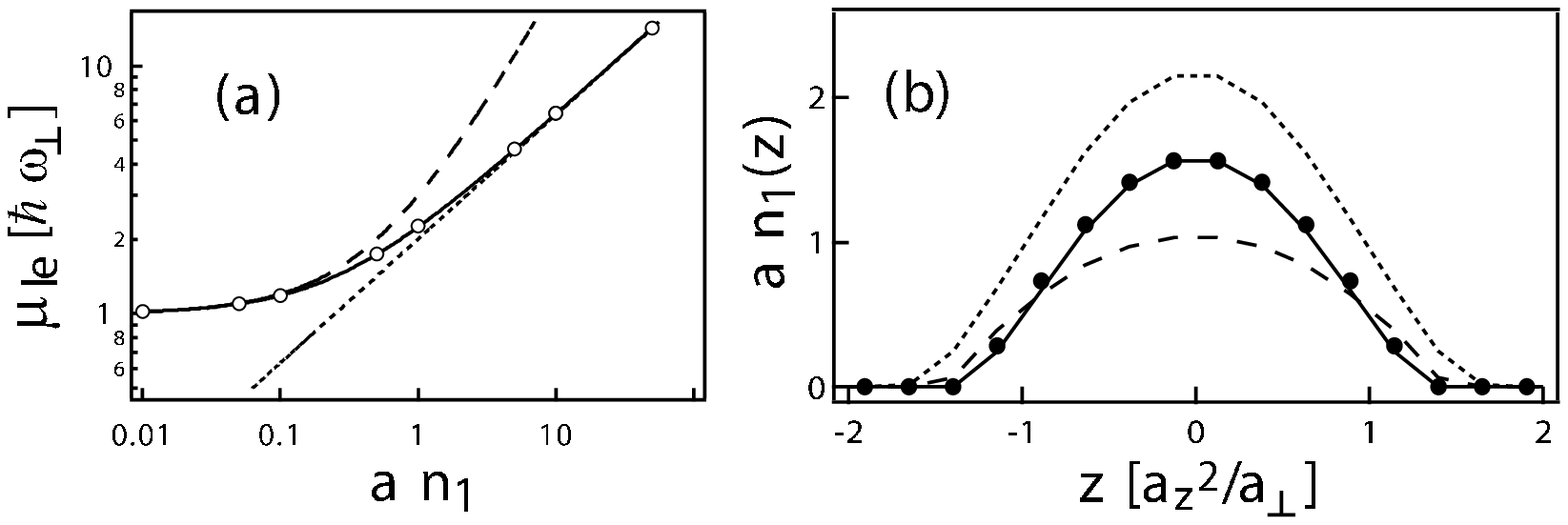}
%\onefigure{fig1_crossover.eps}

\caption{Accuracy of the local density approximation. {\bf (a)}
Local chemical potential near the 3D-1D
crossover, as a function of the local 1D density $n_1$. The
circles show the results of a numerical calculation
\cite{menotti2002a}, undistinguishable at the scale of the figure
from Eq. (\ref{mule}) [solid line]. The dotted and dashed lines show the
3D and 1D Thomas-Fermi limiting cases. {\bf (b)} Integrated density
profiles taking the (harmonic) axial trapping potential into
account. To highlight the crossover, the parameter $\chi=1$ has been chosen
(corresponding to $\mu \approx1.85\hbar\omega_\perp$). The circles,
resulting from a numerical solution of the Gross-Pitaevskii
equation, are indistinguishable from the LDA result
(solid line) at the scale of the figure. The dotted and dashed lines give the
3D and 1D Thomas-Fermi profiles, extrapolated to $\chi=1$ for
comparison.} \label{fig1}
\end{figure}

To obtain this equation of state, it is sufficient to solve (\ref{gp2d}) locally, {\it
i.e.} for each value of $n_1(z)$, or equivalently by considering a
geometry with radial harmonic trapping, but homogeneous 1d density
$n_1=N/2L$, where $2L$ is the axial length of the system (``cylinder model'', see inset in Fig. \ref{fig1}a).
The equation of state in this model was found in
\cite{menotti2002a} by numerical integration. Here we follow an
alternative route, and derive the equilibrium properties by taking
for $f_\perp[n_1]$ a Gaussian trial wavefunction,  whose width
$w_\perp[n_1]$ is a variational parameter, and by minimizing the
{\it chemical potential} $\mu_{\rm l.e.}$. This simple calculation
yields the optimized width $w_\perp[n_1]=a_\perp(1+4 a
n_1)^{1/4}$, where $a_\perp=\sqrt{\hbar/M\omega_\perp}$ is the
radial oscillator length, and the local chemical potential
\begin{equation}\label{mule}
\mu_{\rm l.e.}[n_1]=\hbar\omega_\perp\sqrt{1+4an_1}.
\end{equation}
The very good agreement of Eq. (\ref{mule}) with the available
numerical results \cite{menotti2002a}, illustrated in Fig.
\ref{fig1}a, has been pointed out in \cite{fuchs2003a} on a
phenomenological basis. Here we show that this expression follows
from the condition that $\mu_{\rm l.e.}$ is the lowest eigenvalue of the
Gross-Pitaevskii equation. Note that this approximation does not
correspond to a separability {\it ansatz}, as $n_1$ is in general a
function of the axial coordinate $z$. In this respect, the method
used here differs from the work reported in \cite{das2002a}.

We now reintroduce the axial trapping potential, assuming for
definiteness a harmonic form, $V(z)=\frac{1}{2}M\omega_z^2 z^2$.
From the local equilibrium condition (\ref{lda}) and the equation
of state (\ref{mule}), one finds the density profile of the
trapped gas,
\begin{eqnarray}\label{n1}
n_1(z) & = &
\frac{1}{16a}\frac{V(L)-V(z)}{\hbar\omega_\perp}\left[
\frac{V(L)-V(z)}{\hbar\omega_\perp}+1 \right]  =
\frac{\alpha}{4a}(1-\tilde{z}^2)\left[ \alpha(1-\tilde{z}^2)+4
\right].
\end{eqnarray}
Here we have introduced the parameter
$\alpha=2(\mu/\hbar\omega_\perp-1)$ and $\tilde{z}=z/L$. The
condensate length $L$ is defined by the relation $\mu_{\rm
l.e.}[n_1(L)=0]+\frac{1}{2}M\omega_z^2L^2=\mu$, and is given
explicitly by
\begin{eqnarray}\label{L}
L=\frac{a_z^2}{a_\perp}\sqrt{\alpha},
\end{eqnarray}
where $a_z=\sqrt{\hbar/M\omega_z}$ is the axial oscillator length.
Using $\int_{-L}^L n_{1}(z)dz=N$, and the density profile
(\ref{n1}), we obtain the equation for the key quantity $\alpha$,
\begin{eqnarray}\label{alpha}
\alpha^3(\alpha+5)^2=(15 \chi)^2.
\end{eqnarray}
The only parameter of the calculation, $\chi=N a a_\perp/a_z^2$,
roughly gives the ratio of the interaction energy to the radial
zero-point energy \cite{menotti2002a}. Numerical solution of
Eq. (\ref{alpha}) is straightforward, and obtains the static
properties of the condensate at any confinement strength. In the
limit $\chi\gg5$, the mean-field interaction dominate over the
transverse confinement, and one recovers the well-known 3D TF
result, $\alpha\approx\alpha_{\rm 3D} =(15\chi)^{2/5}$.
Conversely, if $\chi \ll 5$, the transverse motion is almost
frozen and one finds $\alpha\approx\alpha_{\rm 1D}=(3\chi)^{2/3}$.
The crossover between the two regimes occurs approximately for
$\alpha_{\rm 1D}=\alpha_{\rm 3D}$, giving a cross-over value
$\chi_{\rm cross}=5^{3/2}/3\approx3.73$. Our analytical results
are very accurate even in the crossover region, as shown in Fig.
\ref{fig1}b where we compare Eq. (\ref{n1}) to a direct numerical
solution of the GP equation.

Equations (\ref{n1},\ref{L},\ref{alpha}) are the key results of
this paper. The calculation of the density profile for any
strength of the confinement in the transverse direction allows to
deduce a number of interesting quantities. As an example, we focus
in the remainder of the paper on phase fluctuations, a key
phenomenon to understand the physics of 1D gases. In general,
phase fluctuations originate from the very large population of the
1D excited states of the system, {\it i.e.} those in the
low-energy energy range $\hbar\omega_z<\epsilon \ll {\rm min} \{ \mu,\hbar\omega_\perp \}$
(referred to in the following as the ``axial branch'' of excitations). Keeping with the local density approach, we analyze first these
excitations in the cylinder geometry. The axial branch corresponds to
excitations characterized by an axial wavenumber $k$, and no
radial nodes \cite{stringari1998a}. Well below the degeneracy temperature \cite{Td}, the elementary excitations of the phase-fluctuating ensemble obey the same Bogoliubov-De Gennes equations than in the usual coherent
ensemble \cite{petrov2000a}. We introduce the operators $\hat{\phi}$ and $\delta\hat{n}$
describing fluctuations of the phase and of the density, and their
expansion on the set of axial plane waves, $\delta \hat{n}(z)  =
\sum_k \delta n_{k} \mathcal{A}_k(\rho) e^{i k z }
\hat{b}_k/\sqrt{2n_1L}~+~{\rm h.c.} $, and $\hat{\phi}(z)  =
-i\sum_k \phi_{k} e^{i k z} \hat{b}_k/\sqrt{2n_1L}~+~{\rm h.c.}$.
The function $\mathcal{A}_k(\rho)$ describes the radial dependance
of density fluctuations, and the operators $\hat{b}_k$,
$\hat{b}_k^\dagger$ destroy and create one quasi-particle with
wavevector $k$. The Fourier components of $\hat{\phi}$ and
$\delta\hat{n}$ read
\begin{eqnarray}\label{continuitee3}
 \hbar \omega_k^{\rm B} \delta n_k \mathcal{A}_k(\rho)& = & \frac{\hbar k^2}{M} n_0(\rho) \phi_k ,
 \\ \label{Euler3}
 \hbar\omega_k^{\rm B} \phi_k n_0(\rho) & = &
\left(\frac{\hbar^2 k^2}{4M}+U n_0(\rho)\right) \delta n_k
\mathcal{A}_k(\rho)  -\frac{\hbar^{2}\delta n_k}{4 M} {\bf
\nabla_\perp} \left[ n_0(\rho) {\bf \nabla_\perp}
\left(\frac{\mathcal{A}_k(\rho)}{n_0(\rho)}\right) \right].
\end{eqnarray}
For the axial branch of interest, the transverse envelope changes continuously from a flat profile
in the 3D regime to the radial ground state in the 1D regime \cite{stringari1998a}. In
any case, the quantity $\mathcal{A}_k/n_0$ is almost flat; The
corresponding spatial derivatives in (\ref{Euler3}) are strictly zero in the 1D limit, and of order
 $(\hbar\omega_\perp/\mu)^2 \ll 1$ compared to the first term of the right hand side of (\ref{Euler3})
in the 3D limit. We thus neglect their contribution,
and average over transverse degrees of freedom to get rid of the remaining
radial dependance \cite{stringari1998a}. The equations obtained in this way
are of the usual Bogoliubov form, with an excitation spectrum given by
spectrum $\omega_k^{\rm B}=\sqrt{(\hbar k^2/2M)^2 +c_{\rm 1D}^2 k^2}$, and the amplitudes $\delta n_k =n_1 (\omega_{k}/\omega_{k}^{\rm B})^{1/2}$ and $\phi_k  = (\omega_{k}^{\rm B}/4\omega_{k})^{1/2}$. The longitudinal speed of sound is $c_{\rm 1D}^2(k)=U
\overline{n_0\mathcal{A}_k}/M$, and may depend on $k$ through
$\overline{n_0\mathcal{A}_k}=\int d^{(2)}{\bf \rho}~n_0
\mathcal{A}_k$ for $\omega_k^{\rm B} \lesssim \omega_\perp$. This has been suggested in \cite{fedichev2001a} as a
possible mechanism contributing to the decrease of the critical
velocity in 3D elongated gases. The phase coherence properties are however not affected, since they are determined the phonon-like, lowest-energy modes with energy $\lesssim (T/T_\phi)^{1/2} \hbar \omega_z \ll \hbar\omega_\perp$ \cite{petrov2000a,petrov2001a}, for which $\mathcal{A}_k$ is $k$-independent regardless of the transverse regime \cite{stringari1998a}.

\begin{figure}[t]
\includegraphics[width=12cm]{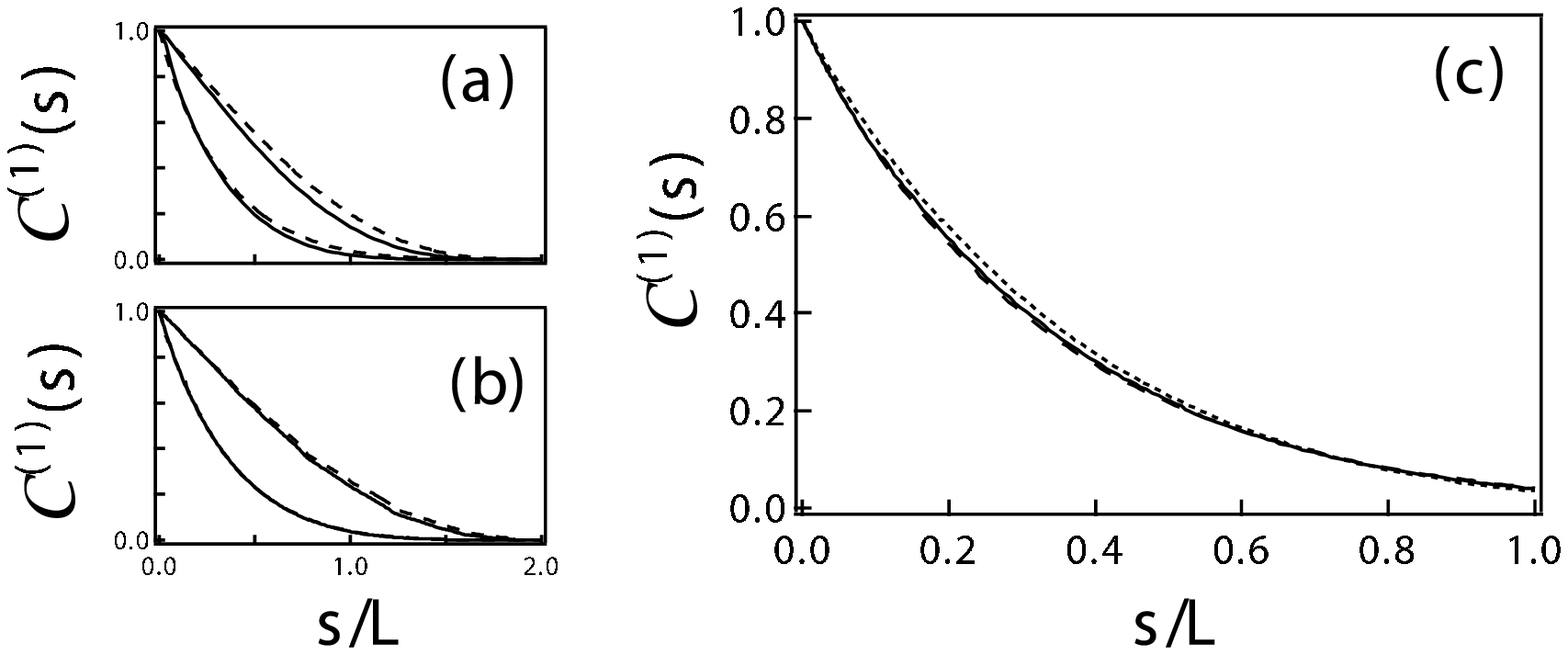}
%\onefigure{fig2_crossover.eps}

\caption{Spatial correlation function of an elongated,
phase-fluctuating condensate for various confinement regimes. In
{\bf (a)}, the spatial correlation function is drawn in the 3D
case as a function of the reduced distance $s/L$, for $T=T_\phi$
(upper curve) and for $T=4T_\phi$ (lower curve). The solid line
follows from a numerical calculation based on the results in
\cite{petrov2001a}, and the dashed line is the LDA. Figure {\bf
(b)} shows the corresponding curves for the 1D case
\cite{petrov2000a}. In {\bf (c)}, the correlation function of an
elongated condensate is plotted for $T=4 T_\phi$ and different
regimes of transverse confinement: $\chi=100$ (3D case, dotted),
$\chi=1$(intermediate case, solid) and $\chi=10^{-2}$ (1D case,
dashed). Despite 4 orders of magnitude of variations in $\chi$,
the functional form of the spatial correlation function is almost unchanged.} \label{fig2}
\end{figure}

We now include the effect of the trapping potential by introducing a local density profile $n_1(z)$ according to
(\ref{lda}) and (\ref{n1}). In the above expressions, one than has to replace everywhere the chemical
potential and density by their local values (see \cite{gerbier2003b} for further details). Using $\langle
\hat{b}_k^\dagger \hat{b}_k\rangle \approx k_{\rm
B}T/\hbar\omega_k^{\rm B}$, the variance of phase fluctuations
then reads
\begin{equation}\label{Dphilda}
\Delta \phi^2(\overline{z},s)=\langle
[\hat{\phi}(z)-\hat{\phi}(z')]^2 \rangle \approx
\frac{T}{T_{\phi}}\frac{n_1(0)}{n_1(\overline{z})}\frac{\mid s
\mid}{L}=
\frac{n_1(0)}{n_1(\overline{z})}\frac{\mid s
\mid}{d_\phi},
\end{equation}
with the relative distance $s=z-z'$ and the mean coordinate
$\overline{z}=(z+z')/2$. The phase temperature in (\ref{Dphilda}) is $k_{\rm B} T_\phi=\hbar^2 n_1(0)/M
L \propto N(\hbar\omega_z)^2/\mu$ \cite{petrov2000a,petrov2001a}, and the phase coherence length is $d_\phi=L T_\phi/T=\hbar^2 n_1(0)/M k_{\rm B} T$. Together with the density envelope (\ref{n1}), the expression
(\ref{Dphilda}) is sufficient to find the long wavelength behavior of the spatial correlation
function \cite{petrov2000a},
\begin{eqnarray} \label{c1lda}
\mathcal{C}^{(1)}\left(\frac{s}{L}\right)& = &
\frac{1}{N}\int_{-L}^{L}
d\overline{z}~\sqrt{n_{1}(\overline{z}+s/2)n_{1}(\overline{z}-s/2)}
\exp(-\frac{1}{2}\Delta \phi^2(\overline{z},s)).
\end{eqnarray}
This expression differs slightly from the one used in
\cite{gerbier2003b} in the treatment of the overlap term, defined
here as $\sqrt{n_{1}(\overline{z}+s/2)n_{1}(\overline{z}-s/2)}$.
The way we write it here yields better agreement with the
numerical calculation of the correlation function in the 3D case
(see Fig.\ref{fig2}b).

Several comments can be made on this expression. First, already
for $T=4T_\phi$, the LDA expression (\ref{c1lda}) agrees well with
the numerical calculation of the correlation function based on
Refs. \cite{petrov2000a,petrov2001a}. This is shown in
Fig.\ref{fig2}a for the 3D case, and Fig.\ref{fig2}b for the 1D
case.  Second, the expression (\ref{c1lda}), which depends on the
dimensionless space variable $s/L$, is a universal function
completely determined by two dimensionless parameters, $\chi$,
which controls the regime of transverse confinement and the
functional form of the density profile, and $T/T_\phi$, which
controls the magnitude of phase fluctuations. The third and most
important conclusion is that the resulting correlation function is
insensitive to a large extent to the transverse regime of
confinement (in other words, to the value of $\chi$). This is
illustrated in Fig.\ref{fig2}b, where we plot the correlation
function for $T=4T_\phi$ and $\chi=100$ (dotted), $\chi=1$ (solid)
and $\chi=10^{-2}$ (dashed). Despite the dissimilar transverse
profiles, which correspond to very different experimental systems
(see Table \ref{t.1}), the axial correlation function are almost
identical, showing almost exponential decay on a $1/e$ length
scale $\approx 1.54 d_\phi$. This very weak dependance on $\chi$
points out to the almost universal nature of thermal phase
fluctuations in ultracold, very elongated trapped gases.
Modifications of the functional profile
$n_1(\overline{z})/n_1(0)$, explicitly present in Eq.
(\ref{c1lda}), are not significant. Rather, the effects of
transverse confinement are almost entirely contained in the
scaling variable $d_\phi \propto n_1(0)$.

The approximate scaling identified here relies on the
zero-temperature equation of state, Eq. (\ref{mule}). Its validity
requires that (i) axial density fluctuations are negligible, and
(ii) that when going to the 3D regime, the presence of a normal
cloud, mostly composed of 3D excited states with energy $\gg
\hbar\omega_\perp$, do not modify significantly the density
profile of the quasicondensate. For (i) to be true, one requires
that $T \ll T_{\rm d}$, where $T_{\rm d}=N\hbar \omega_z$ is the
1D degeneracy temperature \cite{petrov2000a,petrov2001a}. Note
this is always the case in the 3D case since $T_{\rm d} \gg T_{\rm
c}^{(3D)}$ \cite{Td}. To check point (ii), we note that although
the quasicondensate is significantly depopulated for temperature
$T \gtrsim 0.5 ~ T_{\rm c}^{(3D)}$, the mechanical effect on the
density profile is only noticeable for $T\gtrsim 0.8 ~ T_{\rm
c}^{(3D)}$ \cite{gerbier2004b}. Coherence properties of the
strongly degenerate part of the cloud are still described by Eq.
(\ref{c1lda}), and provided the correction to $n_1(0)$ are taken
into account, we expect that the scaling behavior still holds to a
good approximation, since the correlation function is largely
insensitive to the precise functional form of $n_1(z)$.

In conclusion, we have investigated in this paper the crossover
from a very elongated, 3D Bose gas to a 1D situation where
transverse motion is frozen in the limit of vanishing density
fluctuations, {\it i.e.} for a gas strongly in the degenerate
regime. By relying on a local density approximation, we have been
able to compute the radially integrated density profile for any
transverse confinement; we believe these results are simple enough
to prove useful for the analysis of time of flight images of very
elongated samples, with $\epsilon_{\rm int}\sim\hbar\omega_\perp$.
We have applied the method to the problem of phase fluctuations
arising in such geometry at finite temperatures, and have found an
almost ``universal'' behavior of quasicondensates in an elongated
geometry, related to the essentially classical nature of thermal
phase fluctuations.

\begin{acknowledgments}
We acknowledge stimulating discussions with the members
of the Atom Optics group at Institut d'Optique, in particular with
Simon Richard and Isabelle Bouchoule, and thank Joseph H.
Thywissen and Gora Shlyapnikov for critical comments on this
manuscript.
\end{acknowledgments}
%
%%%%%%%%%%%%%%%%%%%%%%%%%%%%%%%%% body %%%%%%%%%%%%%%%%%%%%%%%%%%%%%%%%%
%

% Create the reference section using BibTeX:
%\bibliography{crossover}

\begin{thebibliography}{0}

\bibitem{gorlitz2001a}
\Name{A. G\"orlitz {\it et al.}}
\REVIEW{Phys. Rev.
Lett.}{87}{130402}{2001}.

\bibitem{schreck2001a}
\Name{F. Schreck {\it et al.}} \REVIEW{Phys. Rev.
Lett.}{87}{080403}{2001}.

\bibitem{greiner2001a}
\Name{M. Greiner, I. Bloch, O. Mandel, T. W. Haensch, T.
Esslinger} \REVIEW{Phys. Rev. Lett.}{87}{160405}{2001}.

\bibitem{moritz2003a}
\Name{H. Moritz, T. St{\"o}ferle, M. K{\"o}hl, T. Esslinger}
\REVIEW{Phys. Rev. Lett.}{91}{250402}{2003}.

\bibitem{petrov2000a}
\Name{D. S. Petrov, G.~V. Shlyapnikov, J. T. M. Walraven}
\REVIEW{Phys. Rev. Lett.}{85}{3745}{2000}.

\bibitem{tonks}
\Name{M. Olshanii} \REVIEW{Phys. Rev. Lett.}{81}{938}{1998};
\Name{M. D. Girardeau, E. W. Wright} \REVIEW{Phys. Rev.
Lett.}{84}{5239}{2000}.

\bibitem{qbec}
\Name{J. O. Andersen, U. Al-Khawaja, H. T. C. Stoof}
\REVIEW{Phys. Rev. Lett.}{88}{070407}{2002}; \Name{U. Al-Khawaja,
J. O. Andersen, N. P. Proukakis, H. T. C. Stoof}
\REVIEW{Phys. Rev. A}{66}{013615}{2002};
\Name{C. Mora, Y. Castin} \REVIEW{Phys. Rev.
A}{67}{053615}{2003};
\Name{D. Luxat, A. Griffin} \REVIEW{Phys. Rev.
A}{67}{043603}{2003}; \Name{N. M. Bogoliubov {\it et al.}}\REVIEW{Phys. Rev. A}{69}{023619}{2004}.

\bibitem{petrov2001a}
\Name{D. S. Petrov, G.~V. Shlyapnikov, J. T. M. Walraven}
\REVIEW{Phys. Rev. Lett.}{87}{050404}{2001}.

\bibitem{stringari1998a}
\Name{S. Stringari} \REVIEW{Phys. Rev. A}{58}{2385}{1998}.

\bibitem{dettmer2001a}
\Name{S. Dettmer {\it et al.}} \REVIEW{Phys. Rev.
Lett.}{87}{160406}{2001}.

\bibitem{schvarchuck2002a}
\Name{I. Schvarchuck {\it el al.}} \REVIEW{Phys. Rev.
Lett.}{89}{270404}{2002}.

\bibitem{richard2003a}
\Name{S. Richard {\it et al.}} \REVIEW{Phys. Rev.
Lett.}{91}{010405}{2003}.

\bibitem{gerbier2003b}
\Name{F. Gerbier {\it et al.}} \REVIEW{Phys. Rev.
A}{67}{051602(R)}{2003}.

\bibitem{hellweg2003a}
\Name{D. Hellweg {\it et al.}} \REVIEW{Phys. Rev.
Lett.}{91}{010406}{2003}; \Name{L. Cacciapuoti {\it et al.}} \REVIEW{Phys. Rev.
A}{69}{023619}{2003}.

\bibitem{hansel2001a}
\Name{W. H{\"a}nsel, P. Hommelhoff, T. W. H{\"a}nsch , J.
Reichel} \REVIEW{Nature}{413}{498}{2001}.

\bibitem{ott2001a}
\Name{H. Ott, J. Fort{\`a}gh, G. Schlotterbeck, A. Grossmann,
C. Zimmermann} \REVIEW{Phys. Rev. Lett.}{87}{230401}{2001}.

\bibitem{das2002a}
\Name{K.~K. Das, M. D. Girardeau, E. M. Wright } \REVIEW{Phys. Rev. Lett.}{89}{110402}{2002}.

\bibitem{menotti2002a}
\Name{C. Menotti, S. Stringari} \REVIEW{Phys. Rev.
A}{66}{043610}{2002}.

\bibitem{das2002b}
\Name{K.~K. Das}\REVIEW{Phys. Rev. A}{66}{053612}{2002}.

\bibitem{fuchs2003a}
\Name{J.-N. Fuchs, X. Leyronas, R. Combescot} \REVIEW{Phys.
Rev. A}{68}{043610}{2003}.

\bibitem{alkhawaja2004a}
\Name{U. Al Khawaja, N. P. Proukakis, J. O. Andersen, M. W. J.
Romans, and H. T. C. Stoof} \REVIEW{Phys. Rev.
A}{68}{043603}{2003}.

\bibitem{cctcargese}
\Name{C.\ Cohen-Tannoudji, C.\ Robilliard} \Review{C.\ R.\
Acad.\ Sci.\ Paris, t.2, s{\'e}rie IV}{445}{2001}.

\bibitem{Td}
The degeneracy temperature in the 3D case is just the critical temperature, $T_{\rm c}^{(3D)} \approx 0.94 \hbar \overline{\omega}N^{1/3} $, where the mean trapping frequency is $\overline{\omega}=(\omega_\perp \omega_z)^{1/3}$. In 1D, it is given instead by $T_{\rm d} = N \hbar \omega_z$. The  parameter $\eta=N \omega_z/\omega_\perp$ fixes the ratio $T_{\rm c}/\hbar \omega_\perp \sim \eta^{1/3}$ and  $T_{\rm c}^{(3D)}/T_{\rm d} \sim \eta^{-2/3}$.

\bibitem{fedichev2001a}
\Name{P. O. Fedichev, G. V. Shlyapnikov} \REVIEW{Phys. Rev.
A}{63}{045601}{2001}.

\bibitem{gerbier2004b}
\Name{F. Gerbier {\it et al.}} \REVIEW{to appear in Phys. Rev.
A}{2004}.



\begin{table}
\caption{Realistic experimental parameters illustrating the various regimes considered in the paper. Notations are explained in the text.}
\label{t.1}
\begin{center}
\begin{tabular}{lccccccccr}
\hline    & Atom number & $\omega_\perp/2\pi$ &  $\omega_z/2\pi$  &  $\chi$ & $T_{\rm \phi}$ & $T_{c}^{(3D)}$& $T_{\rm d}$\\
\hline  3D regime: & $10^6$ & $1$ kHz & $20$ Hz & 90 & 120 nK & 1.2 $\mu$K & 1 mK  \\
$^{23}$Na, optical trap & & & & & & &\\
\hline Crossover regime: & $5\times 10^4$ & $760$ Hz & $5$ Hz & 3 & 35 nK & 250 nK & 120 $\mu$K \\
$^{87}$Rb, magnetic trap \cite{richard2003a} & & & & & & &\\
\hline 1D regime: & $200$ & $20$ kHz & $50$ Hz & 0.02 & 4 nK & 720 nK & 480 nK \\
$^{87}$Rb, 2D optical lattice \cite{greiner2001a} & & & & & & &\\
\hline
\end{tabular}
\end{center}
\end{table}



\end{thebibliography}
%\bibliographystyle{plain}

\end{document}